\documentclass{svmult}
\usepackage{type1cm}
\usepackage{makeidx}
\usepackage{graphicx}
\usepackage{multicol}
\usepackage[bottom]{footmisc}
\usepackage{newtxtext}
\usepackage{newtxmath}
\makeindex
\usepackage[ruled,vlined,linesnumbered]{algorithm2e}
\usepackage{caption}
\usepackage{url}

\begin{document}

\title*{Using a Bi-directional LSTM Model with Attention Mechanism trained on MIDI Data for Generating Unique Music}
\titlerunning{Bi-directional LSTM with Attention for Generating Unique Music}
\author{Ashish Ranjan, Varun Nagesh Jolly Behera and Motahar Reza}
\institute{Ashish Ranjan \at R\&D Division, Just Another Media Laboratory (JAM Lab), Mumbai
\newline\email{ashish.ranjan@jamlab.in}
\and Varun Nagesh Jolly Behera \at R\&D Division, Just Another Media Laboratory (JAM Lab), Mumbai
\newline\email{varun.behera@jamlab.in}
\and Motahar Reza \at Department of Mathematics, GITAM University Hyderabad Campus, Hyderabad
\newline\email{mreza@gitam.edu}}
%
%
\maketitle

\abstract{Generating music is an interesting and challenging problem in the field of machine learning. Mimicking human creativity has been popular in recent years, especially in the field of computer vision and image processing. With the advent of GANs, it is possible to generate new similar images, based on trained data. But this cannot be done for music similarly, as music has an extra temporal dimension. So it is necessary to understand how music is represented in digital form. When building models that perform this generative task, the learning and generation part is done in some high-level representation such as MIDI (Musical Instrument Digital Interface) or scores. This paper proposes a bi-directional LSTM (Long short-term memory) model with attention mechanism capable of generating similar type of music based on MIDI data. The music generated by the model follows the theme/style of the music the model is trained on. Also, due to the nature of MIDI, the tempo, instrument, and other parameters can be defined, and changed, post generation.}

\keywords{LSTM, attention, MIDI, music}

\section{Introduction}
"Beauty is in the eye of the beholder", is a common phrase used in English. This phrase becomes more true when we consider something like art \cite{kumar2010beauty}. Art is something very human by nature. It is highly subjective, entirely based on the observer and their experiences \cite{johnston1993beauty}. For example, it is hard for some people to accept rap as a form of music, and as a form of art \cite{shushterman1991rap}. Art always has a pattern that is recognizable, while still containing some irregularities \cite{field2009art}. This mixture of patters and irregularities are able to please us and we start to like that form of art.

Human beings, for the longest time, have believed that art is the forefront where a machine may never beat them \cite{varshney2013cognition}. But, with the advent of powerful computers and state of the art machine learning algorithms, there might be more prominent AI artists in the future \cite{besold2015creativity}. Since machines don't need rest, and an AI could generate 1000s of art pieces per hour, there will be increased efficiency and cost-effectiveness in art based sectors \cite{toivonen2020creativity}. Generative Adversial Networks (GAN) \cite{goodfellow2014generative}, are very popular these days, especially for creating hyper-realistic faces \cite{gauthier2014conditional}. These systems have huge face datasets to learn from \cite{gao2007face}\cite{kasinki2008face}\cite{ma2015face}. An example of such a system is DeepFake \cite{korshunova2017face}, which superimposes the generated face over a subject’s face in an image or video so that it seems like some other person is present in the image or video. Another example is Neural Style Transfer \cite{li2017style}, which learns the art style from one image and applies the style to another image effectively transferring the style. This technique has been used to recreate the art style of famous painters \cite{gatys2016style}.

A more technical term for machines creating art, is Computational Creativity \cite{colton2012creativity}. It is an interdisciplinary field spanning topics like AI, cognition, arts, design and philosophy \cite{minsky1967creativity}. It is the synthesis of art, in a partially or fully automated way. This field of study focuses on creating algorithms that mimic human creativity, understanding human creativity, mathematically defining creativity, or create systems that assist and enhance the human creative process \cite{gero2000creativity}. Music is probably one of the toughest computational creativity problems, as it is difficult to analyze by a computer \cite{leach1995nature}. It is easier to analyze an image, as there is no temporal component, and the whole image can be observed at once. It's local and global dependencies can be easily analyzed. Whereas, in the case of music, the temporal dimension, and the fact that it is highly coherent and rhythmic, poses even more problems for the computer. The task of generating melodic audio which sound realistic is a tough job and is an open problem in the field of machine learning \cite{papadopoulos1999composition}. Imagine a DJ who does not only mix music but also composes them on the fly. Another example is generating music based on the current mood of the user or even replicate the style of famous musicians and generate new music. The possibilities are endless.

This paper explores a methodology for generating unique music, that is melodic, using deep learning. The methodology involves a bi-directional LSTM with attention mechanism. This system learns the melodic and rhythmic patterns and dependencies in order to generate novel music as output. MIDI \cite{moog1986midi}, a popular high level representation for music has been used.

\section{Algorithmic Composition}
Algorithmic composition is the process of using some rules and combining parts of music to form a whole \cite{nierhaus2009composition}. It is not necessarily computer based, an example being Mozart's musical dice game \cite{hedges1978dice}.

Algorithmic composition has been around for centuries, with Pythagoras suggesting Music and Mathematics being the same \cite{fauvel2006pythagoras}. One of the first computational models used for generating music was using Markov chains \cite{mcalpine1999music}. Other older techniques include species counterpoint, d’Arezzo’s Micrologus \cite{sullivan1989micrologus}, Schoenberg’s twelve-tone technique \cite{dahlaus1987music}, or Cage’s aleatoric music \cite{jeongwon2002aleatoric}. Algorithmic composition is not restricted to just computational means, as seen by the above mentioned methods, but that is not the focus point of this paper. The focus will be on current AI techniques, and more specifically, NLP techniques.
 
Some other algorithms, including fractals \cite{hsu1991music}, statistical models \cite{conklin2003music}, L-systems \cite{worth2005music}, and even random real world data, can in fact be used as a basis for music generations. This is because all of them have some sort of mathematical patterns. Generally algorithmic composition is either music composed by a machine, or music created with the help of a machine.

Using the most prominent features, algorithmic composition may be categorised as
\begin{enumerate}
    \item \textbf{Mathematical Models}: These refer to stochastic processes, and more specifically Markov chains, used for real-time algorithmic composition, due to it's low complexity. An example of such a model used for algorithmic composition is Cybernetic Composer \cite{ames1992cyber}.
    \item \textbf{Knowledge based Systems}: These are systems which use symbols and specified rules with constraint to achieve algorithmic composition. An example of this is CHORAL \cite{ebcioglu1988choral}. Building systems like these for music is difficult and time taking.
    \item \textbf{Grammars}: A grammar set for a musical language can be defined, and it's rules used randomly to generate unique music. An Augmented Transition Network (ATN) \cite{woods1970atn} is used to accomplish this task. The music generated may not be of optimal quality, as the semantics defined may not be robust.
    \item \textbf{Evolutionary Methods}: Genetic Algorithms (GA) \cite{whitley1994genetic} are used in this approach. They either use an objective fitness function or a human fitness function. Since music is subjective, the latter may produce undesirable results \cite{horner1991genetic}.
    \item \textbf{Systems which Learn}: These consists of machine learning based systems, especially ANNs \cite{browne2001music}. This paper explores usage of an LSTM based learning system \cite{mangal2019lstm}.
    \item \textbf{Translational Models}: Information from non-musical media is translated to musical media. For example, using time series data generally used for load forecasting and instead restructuring the dimensions as pitch and scale.
    \item \textbf{Hybrid Systems}: Using any of these systems in combination forms a hybrid system. Basically an ensemble model for algorithmic composition.
\end{enumerate}

\section{Related Work}
Generating music is a task that has researchers all around the world producing new ideas. Even in some cases, the main focus is not on music generation but when conditioned on musical dataset, music can be generated. Such is the case with WaveNet \cite{oord2016wavenet}.

“WaveNet” is a probabilistic autoregressive generative model which can generate raw audio. It was developed by Deepmind and has been used in Text-To-Speech applications. The output produced by WaveNet is very realistic and the voices it creates are almost humanlike. Its effectiveness can be summarized by the high Mean Opinion Score (MOS) mentioned in the paper in which it was introduced and the fact that it has been implemented as the voice of Google Assistant. It is not limited to just Text-To-Speech applications as when the developers trained it with musical dataset, realistic sounding rhythmic audio was generated which can easily be identified as music. The parallel version of WaveNet is very fast \cite{oord2018pwavenet}. It can generate 20 seconds of audio in just 1 second. Since it is a Convolutional Neural Network, training it is very easy for the computer as opposed to Recurrent Neural Network which is generally suited to handle sequences of data such as audio.

Another paper, “A Neural Parametric Singing Synthesizer” or NPSS \cite{blaauw2017npss}, uses a modified version of WaveNet to generate music based on features produced by a parametric vocoder. This produces great results which is expected from WaveNet.

Recurrent Neural Networks have also been used for the purpose of music generation. One such application can be seen in the paper “LSTM Based Music Generation System” \cite{mangal2019lstm}. Since audio samples at any instant highly depend on previous samples for coherent audio, Recurrent Neural Network, especially Long Short-term Memory can be used to take care of these dependencies. This approach works well with classical and jazz music because these are more free-form and without definite song structure like verse, chorus and bridge. This is because it only needs to know the recent context of notes to generate the next note.

Generative Adversarial Networks (GANs) can also be used for the purpose of music generation. One such system is “WGANSing” \cite{chandna2019wgan}, a Multi-Voice Singing Voice Synthesizer Based on the Wasserstein-GAN – It is a General Adversarial Network. It employs two networks. One called the generator which generates new music and tries to fool the other network and the other called the discriminator which tries to discriminate between real musical recording and a generated music. After each epoch of training, both the networks get better at their jobs and the generator can be used to generate music.

OpenAI has also worked in this field and created MuseNet. It uses the same general-purpose unsupervised technology as GPT-2 \cite{radford2019gpt}, a large-scale transformer model trained to predict the next token in a sequence, whether audio or text. It is able to extend a music whose starting fragment is given as input or it can generate music from scratch. The live example on their blog shows its performance. There we can select the target style in which the music is to be generated and the starting fragment of a song (which can be from scratch as well).

Some applications such as jukedeck.com (currently offline), aiva.ai, ampermusic.com and many more have commercialized their music generation system and have been deployed as a service to generate custom royalty free music for its users.

Frameworks such as “Magenta” have been made available to the world for researchers to work in the field of generating art by a computer. Magenta in particular, is an open source research project exploring the role of machine learning as a tool in the creative process. It is available for Python and JavaScript, built with TensorFlow. It includes utilities for manipulating source data (primarily music and images), using this data to train machine learning models, and finally generating new content from these models.

Another paper, “The challenge of realistic music generation: modelling raw audio at scale” \cite{dieleman2018music} explores Autoregressive Discrete Autoencoders (ADAs) as a means to enable autoregressive models such as WaveNet to capture long-range correlations in waveforms which otherwise captures local signal structures. The use of autoencoders in this aspect has also been demonstrated by the YouTube channel Code Parade which developed the Neural Composer. It shows the use of Principal Component Analysis to choose the top 40 principal components and use them as control sliders to generate custom music.

\section{The Problem of Music Generation}
Music often brings nostalgia in humans. It is an art form with which we attach emotions. This paper aims to generate realistic music in a sense that it can be recognized as music made by humans. In order to achieve this goal, the problem statement may be divided into several subtasks.

\begin{enumerate}

    \item The system should be able to generate new audio using high-level representation such as MIDI (The first step could be to generate MIDI). 

    \item The system should learn the rhythmic sequences of music during the training phase and produce similar but novel music as output. For audio, each sample depends on the context of the previous samples. This is even more true in case of music. Generally, the context of the music samples depends on the context of all of the previous samples. This is best implemented in WaveNet and made computationally efficient by using Dilated Convolutions which makes large skips in the data to get a larger receptive view of the data.

    \item Based on some input parameters such as the variance control of the principal components in the Neural Composer, the system should be able to produce appropriate results. These components change the output, ever so slightly and when they are changed in combination, newer music is generated.

    \item Overall, the output should be melodic. This is obvious as it is the goal of the paper.

    \item The output should have a decent MOS (Mean Opinion Score, i.e., if people are able to identify that the music is generated by a computer). Since, evaluating audio is difficult as it is highly subjective. The best way to check the performance of a music generation system is to conduct a survey in which, human participants try to guess if a music being played has been generated by a computer or it is an actual human performance.
    
\end{enumerate}

Some of the optional requirements are:

\begin{enumerate}

    \item Inputs such as mood, genre, instrument, mode, key, tempo, etc. could be input parameters. This requires labelled data and can be difficult.

    \item Future objective could be to integrate lyrics generator and vocal synthesizer.

\end{enumerate}

\section{Dataset}

Some of the dataset that have been identified are:

\begin{enumerate}

    \item \textbf{The MAESTRO Dataset}: MAESTRO \cite{hawthorne2018maestro} (MIDI and Audio Edited for Synchronous TRacks and Organization) is a dataset composed of over 200 hours of paired audio and MIDI recordings from ten years of International Piano-e-Competition. Audio and MIDI files are aligned with ~3 millisecond accuracy and sliced to individual musical pieces, which are annotated with composer, title, and year of performance. Uncompressed audio is of CD quality or higher (44.1–48 kHz 16-bit PCM stereo). The labels are not according to our requirement. The size of the dataset in two formats are: WAV: 103 GB (122 GB uncompressed), MIDI: 57 MB (85MB uncompressed).

    \item \textbf{The Million Song Dataset}: The core of the dataset \cite{bertin2011million} is the feature analysis and metadata for one million songs, provided by The Echo Nest. The dataset does not include any audio, only the derived features. Note, however, that sample audio can be fetched from services like 7digital, using code. Songs are tagged (artists, release, album, etc.). It contains more usable labels, such as danceability, key, mode, tempo, etc. for our purpose. The size of the dataset is 280 GB (Full Dataset, i.e., One Million Songs) or 1.8 GB (1\% or 10000 songs).

    \item \textbf{MagnaTagATune Dataset}: This dataset \cite{law2009input} contains the most versatile collection of labels. But the labels are often unorganized and duplicated like a song has been labeled for both classic and classical genre. Extensive preprocessing will be required to work with this dataset. It can be used for genre, instrument and mood classification for music.

    \item \textbf{Creating Custom Dataset}: Adequate dataset for the optional objectives do not exist so a custom dataset has to be created. This requires time and effort for manually labeling the data.

\end{enumerate}

The format of these datasets is either WAV which is directly in the audio domain or MIDI which gives the idea of playing what note at what time. Both of these formats are good but using MIDI is generally easier.

\section{Data Representation}
In order to generate music, it is necessary to understand what the data is. Music is generally represented in the following ways.

\subsection{Notes}
\begin{figure}
\centering
\captionsetup{justification=centering}
\includegraphics[width=\textwidth,height=\textheight,keepaspectratio]{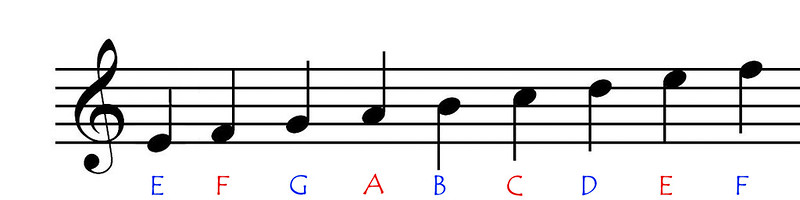}
\caption{Data Representation: Musical Notes}
\label{fig:1}
\end{figure}

A note \cite{strunk1942music} is the symbolic representation of musical sounds. This can be seen in Fig.\ref{fig:1} Notes may be represented in English as A, B, C, D, E, F and G. This totals up to 7 different notes. Another popular representation is Do, Re, Mi, Fa, So, La and Ti. The octave (eighth note), is same as the first, with a double the frequency value, meaning it has a higher pitch. In addition to the note symbol, written notes may also have a note value. This note value determines the notes duration.

The frequency of these notes are measured in hertz (Hz), as mechanical systems produce them. There are 12 notes with fixed frequencies, defined around the central note A$_4$. This note has frequency 440 Hz.

\subsection{Raw Audio}
\begin{figure}
\centering
\captionsetup{justification=centering}
\includegraphics[width=\textwidth,height=\textheight,keepaspectratio]{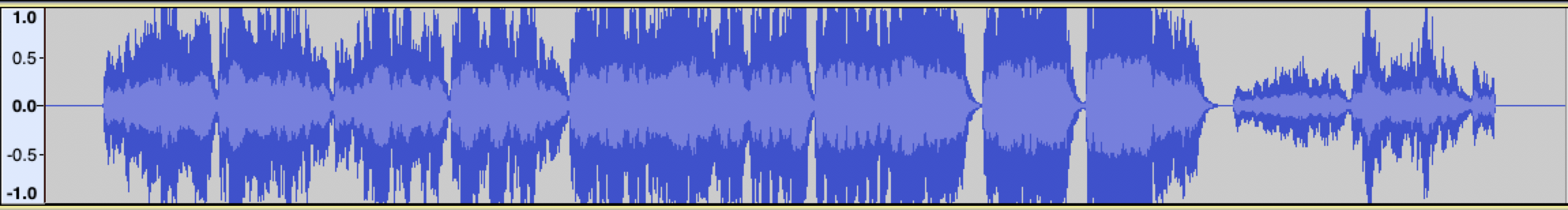}
\caption{Data Representation: Raw Audio}
\label{fig:2}
\end{figure}

Uncompressed audio is stored in raw form is called Raw Audio \cite{tzanetakis2000audio}. Fig.\ref{fig:2} shows the waveform. This file has no header, and hence cannot be played without some user input. So, generally file formats like WAV and AIFF are used, as they are lossless and have nearly the same size. Other lossless formats include FLAC, ALAC and WMA Lossless. Most popular of these are the WAV and FLAC formats.

WAV (Waveform Audio File Format) is an audio file format based on the RIFF bit-stream format. It is similar AIFF as both are based o RIFF. It is typically uncompressed and uses linear pulse-code modulation for bit-stream encoding. Although it is possible to store compressed audio using the WAV format. WAV format is generally used when the best audio quality is required. Generally audio is encoded at 44.1 kHz and 16 bit sample rate.

FLAC (Free Lossless Audio Codec) is a lossless format for storing audio. This codec allows for lossless compression ranging between 50\% to 70\%. It can be easily decoded and streamed,

\subsection{MIDI}
\begin{figure}
\centering
\captionsetup{justification=centering}
\includegraphics[width=\textwidth,height=\textheight,keepaspectratio]{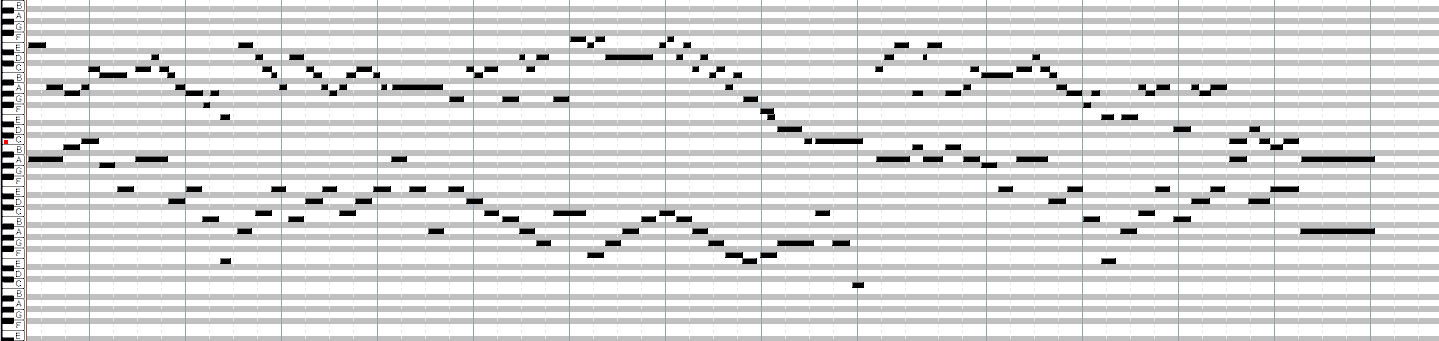}
\caption{Data Representation: MIDI}
\label{fig:3}
\end{figure}

MIDI \cite{moog1986midi} (Musical Instrument Digital Interface) is a universal protocol/interface for many electronic instruments and computers. It allows for ease in recording various instruments and simultaneously interfacing with them via a computer. The file format for this interface is called SMF (Standard MIDI File), and usually has a .mid extension. These files are very small and can be easily transferred between various devices. The header of an SMF contains a lot of essential information like tempo or pitch. Actual audio is not stored in SMF, instead only storing information like pitch and tempo. This is the reason for their extremely small sizes and high portability. This even allowed for licensed MIDI files on FLoppy drives. The main advantage of MIDI is that the instrument type can be easily changed with various software, and that MIDI is great for building the basis of a good soundtrack. Fig.\ref{fig:3} shows the graphical representation of MIDI.

\subsection{Piano Rolls}
\begin{figure}
\centering
\captionsetup{justification=centering}
\includegraphics[width=\textwidth,height=\textheight,keepaspectratio]{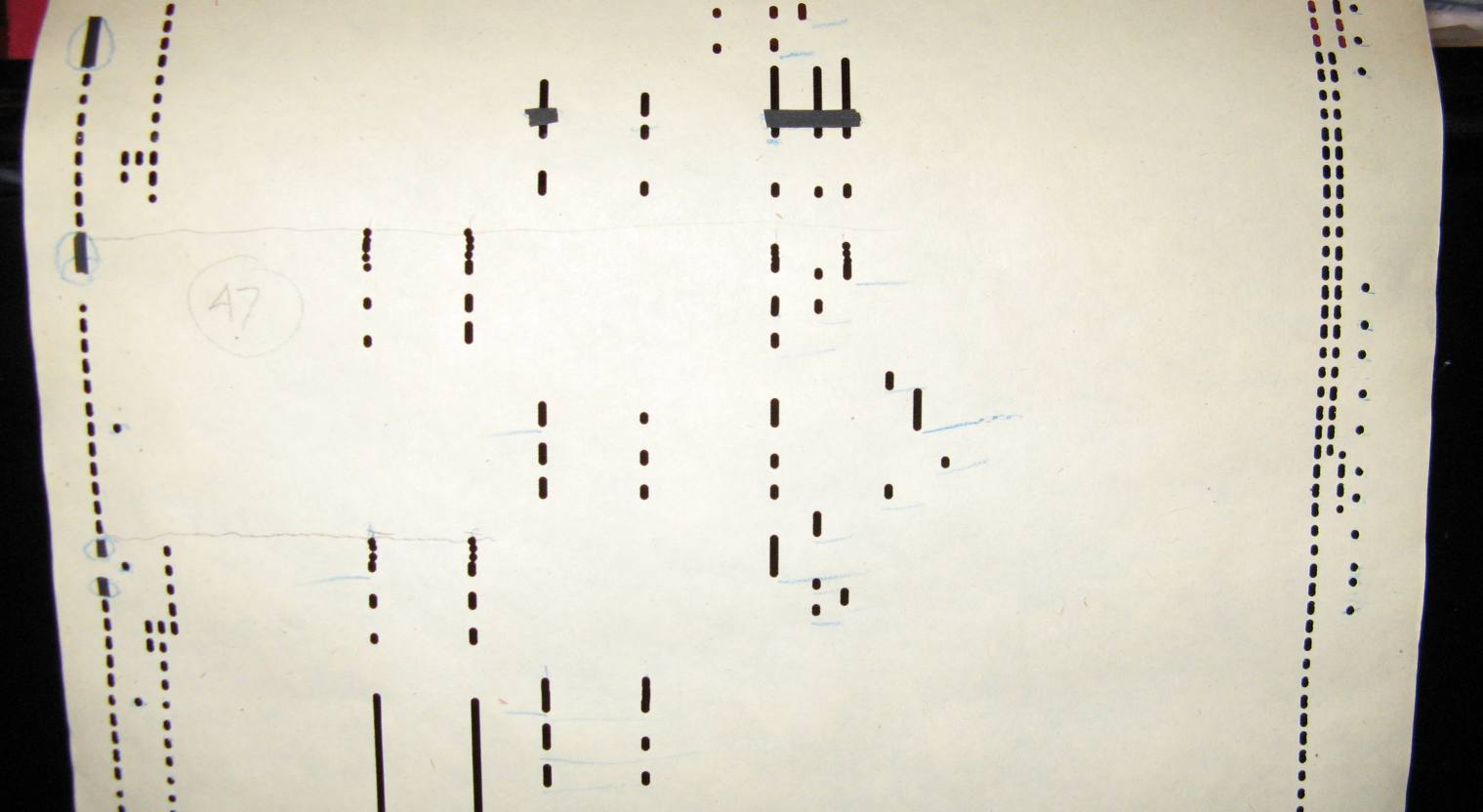}
\caption{Data Representation: Piano Roll}
\label{fig:4}
\end{figure}

Piano Rolls \cite{shi2017piano} are rolls of papers with holes in them as can be seen in Fig.\ref{fig:4} The holes are the data for the notes. They are the logical predecessor to MIDI and an inspiration for it. In softwares, piano rolls generally refer to the visual representation of MIDI data. It is theoretically possible to use this visual data as input for some image based generation model, like GANs. This may be explored in future work.

\section{Data Encoding}
In this paper a custom data set was used, comprising of 92 MIDI files. These are modern video game soundtracks.

\begin{figure}
\centering
\captionsetup{justification=centering}
\includegraphics[width=\textwidth,height=\textheight,keepaspectratio]{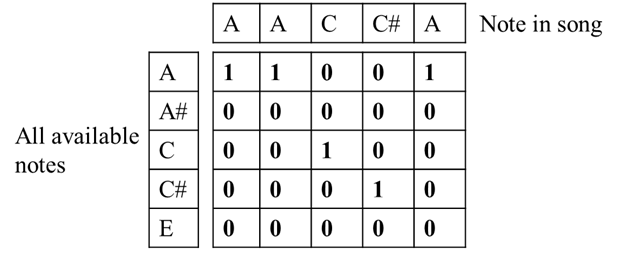}
\caption{Encoding Technique}
\label{fig:5}
\end{figure}

As MIDI files cannot be directly used in an LSTM, they need to be encoded into an appropriate input vector. For this purpose, one hot encoding methodology is used. For each note in the MIDI file, a corresponding true/false value is set. For example, if the note 'A' is present at a specific time step, then the column value corresponding to 'A' will be set to true, and all other set to false. This can be observed in the Fig.\ref{fig:5}, where each column represents an encoded vector for each note present in the song. These generated vectors become the input to the proposed model.

\section{Proposed Model}
A seq2seq \cite{sutskever2014seq} model with attention mechanism \cite{vaswani2017attention} is used for music generation. This is due to seq2seq being able to generate one sequence from another without running into the vanishing gradient problem. It uses an encoder layer and a decoder layer. The encoder layer converts an item into it's corresponding hidden vector and context. The decoder does the opposite of this. In between these two layers is the attention layer. The attention layer helps remember short as well as long term dependencies.

\begin{figure}
\centering
\captionsetup{justification=centering}
\includegraphics[width=\textwidth,height=\textheight,keepaspectratio]{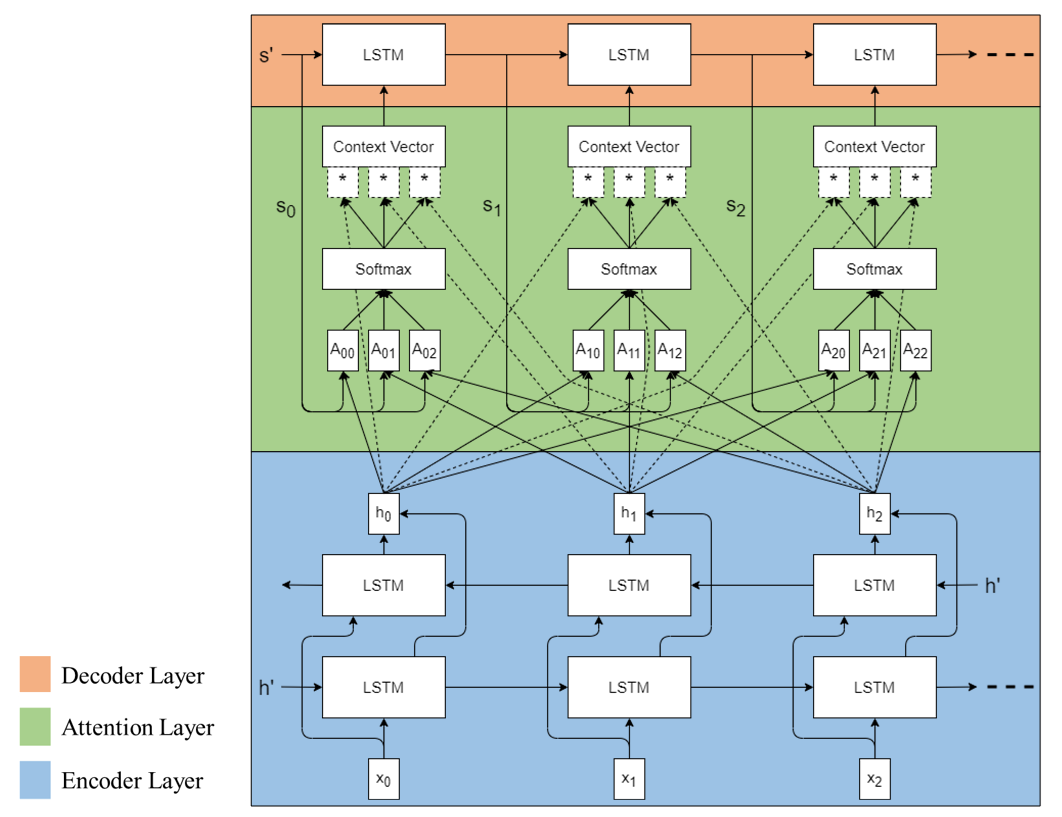}
\caption{Proposed Model}
\label{fig:6}
\end{figure}

\begin{figure}
\centering
\captionsetup{justification=centering}
\includegraphics[width=\textwidth,height=\textheight,keepaspectratio]{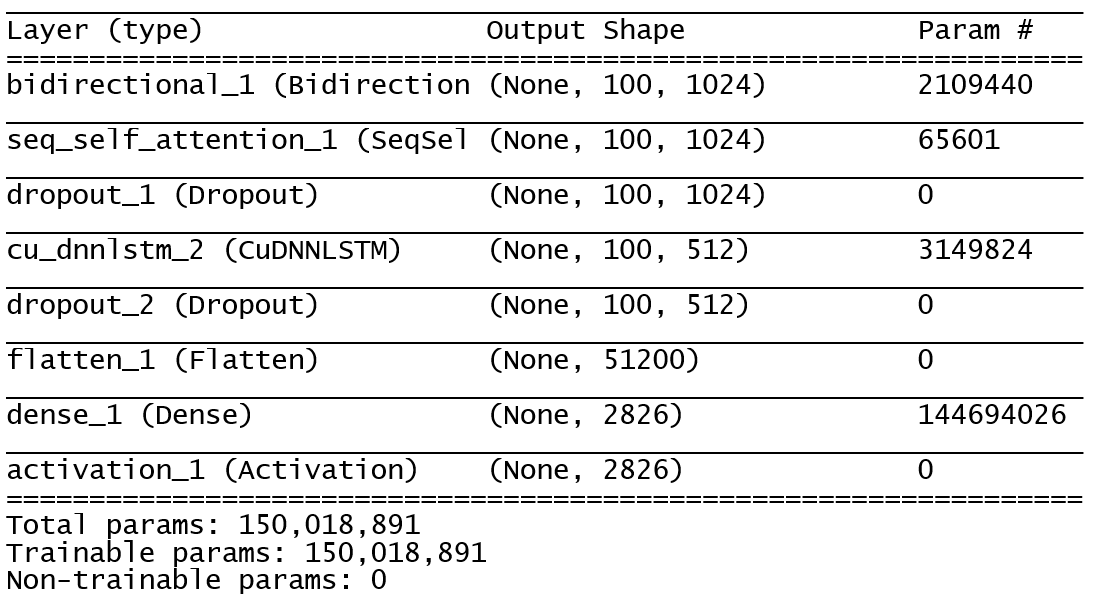}
\caption{Keras Model Specifications}
\label{fig:7}
\end{figure}

The seq2seq model used is an LSTM. The Luong attention mechanism is used with it. The encoder layer consists of a bi-directional LSTM layer and the decoder layer uses a regular LSTM layer. The encoded input vector is converted to hidden states by encoder. This is fed into the attention layer, which produces a context vector. Finally, the context vector thus produced will be used to decode the output sequence The model has been shown in Fig.\ref{fig:6}. It's specifications can be seen in Fig.\ref{fig:7}.

\section{Experimental Results}
The Google Colaboratory platform was used for training the model. This platform provides an Intel(R) Xeon(R) CPU @ 2.20GHz, 12 GB of RAM, and an NVIDIA Tesla K80 GPU.

\begin{figure}
\centering
\captionsetup{justification=centering}
\includegraphics[width=\textwidth,height=\textheight,keepaspectratio]{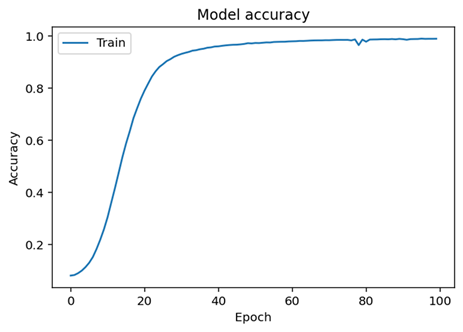}
\caption{Model Accuracy}
\label{fig:10}
\end{figure}

\begin{figure}
\centering
\captionsetup{justification=centering}
\includegraphics[width=\textwidth,height=\textheight,keepaspectratio]{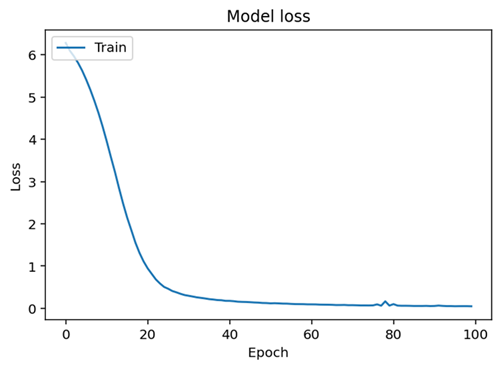}
\caption{Model Loss}
\label{fig:11}
\end{figure}

The training took around 4 hours on this system to achieve an accuracy of 93.02\% at 31 epochs. At this point the training was stopped to avoid overfitting, so as to ensure variety in the generated music. This can be seen in figures \ref{fig:10} and \ref{fig:11}.

\begin{figure}
\centering
\captionsetup{justification=centering}
\includegraphics[width=\textwidth,height=\textheight,keepaspectratio]{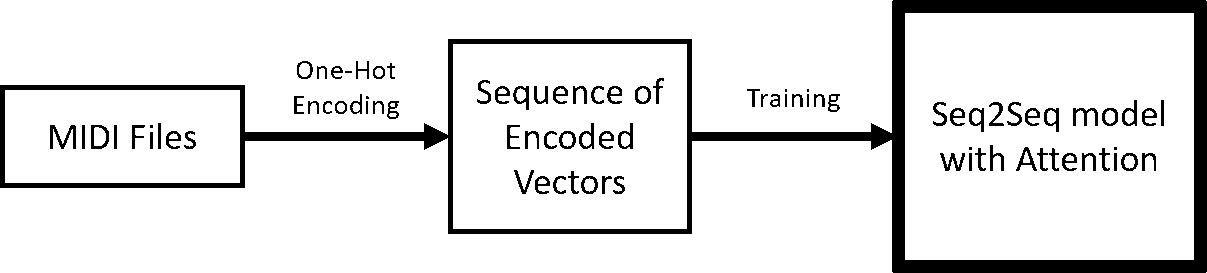}
\caption{Training Workflow}
\label{fig:8}
\end{figure}

The model was trained on a dataset comprising of only 92 MIDI files, which essentially is a small dataset. This was done for faster prototyping. The workflow can be seen in Fig.\ref{fig:8}.

After training, the model can be used for prediction, i.e., music generation. The input and output of the system are as follows.

\begin{enumerate}

    \item \textbf{Input}: A random note or a sequence of notes.
    
    \item \textbf{Output}: The random note or the sequence of notes extended to a length of nearly two minutes. This sequence is rhythmic.
    
\end{enumerate}

\begin{figure}
\centering
\captionsetup{justification=centering}
\includegraphics[width=\textwidth,height=\textheight,keepaspectratio]{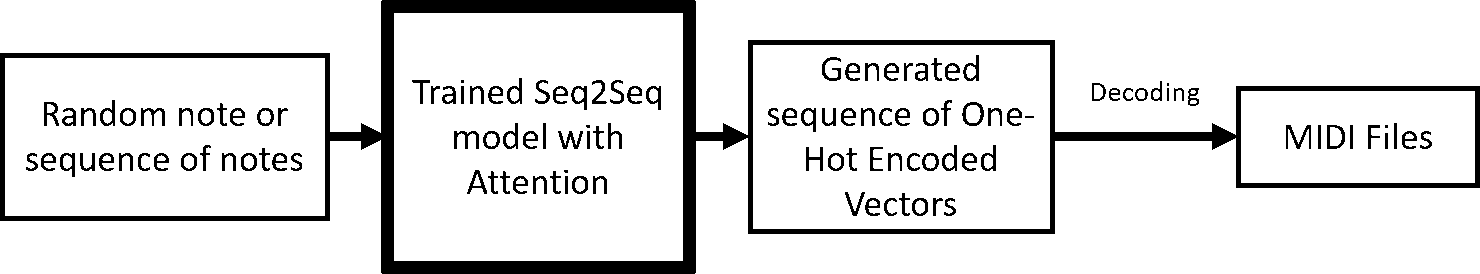}
\caption{Prediction Workflow}
\label{fig:9}
\end{figure}

The prediction workflow is shown in Fig.\ref{fig:9}. The output generation takes approximately 4-5 seconds to complete. This means it is highly efficient.

The model produces a sequence of One-Hot encoded vectors which need to be converted back into MIDI format. To do this, reverse process of the initial One-Hot encoding can be done. Since the produced results are audio files, no image is produced and cannot be shown on paper. The model generates results which sound fairly realistic and the music thus produced was not actually given as input. Though, some nuances of the input songs can be observed. This means that we can tell that the model has learnt from the input songs.

The results can be improved by training on a much larger dataset. In our attempt, we trained on only 92 songs for faster prototyping but this can be extended to huge datasets as well.

A valid question may arise. What if the same starting note is used multiple times? The answer is that it is most likely to produce the same output but this can be solved easily. The input can be extended to a sequence of notes. This sequence of notes can make space for many random combinations of starting notes.

\begin{figure}
\centering
\captionsetup{justification=centering}
\includegraphics[width=\textwidth,height=\textheight,keepaspectratio]{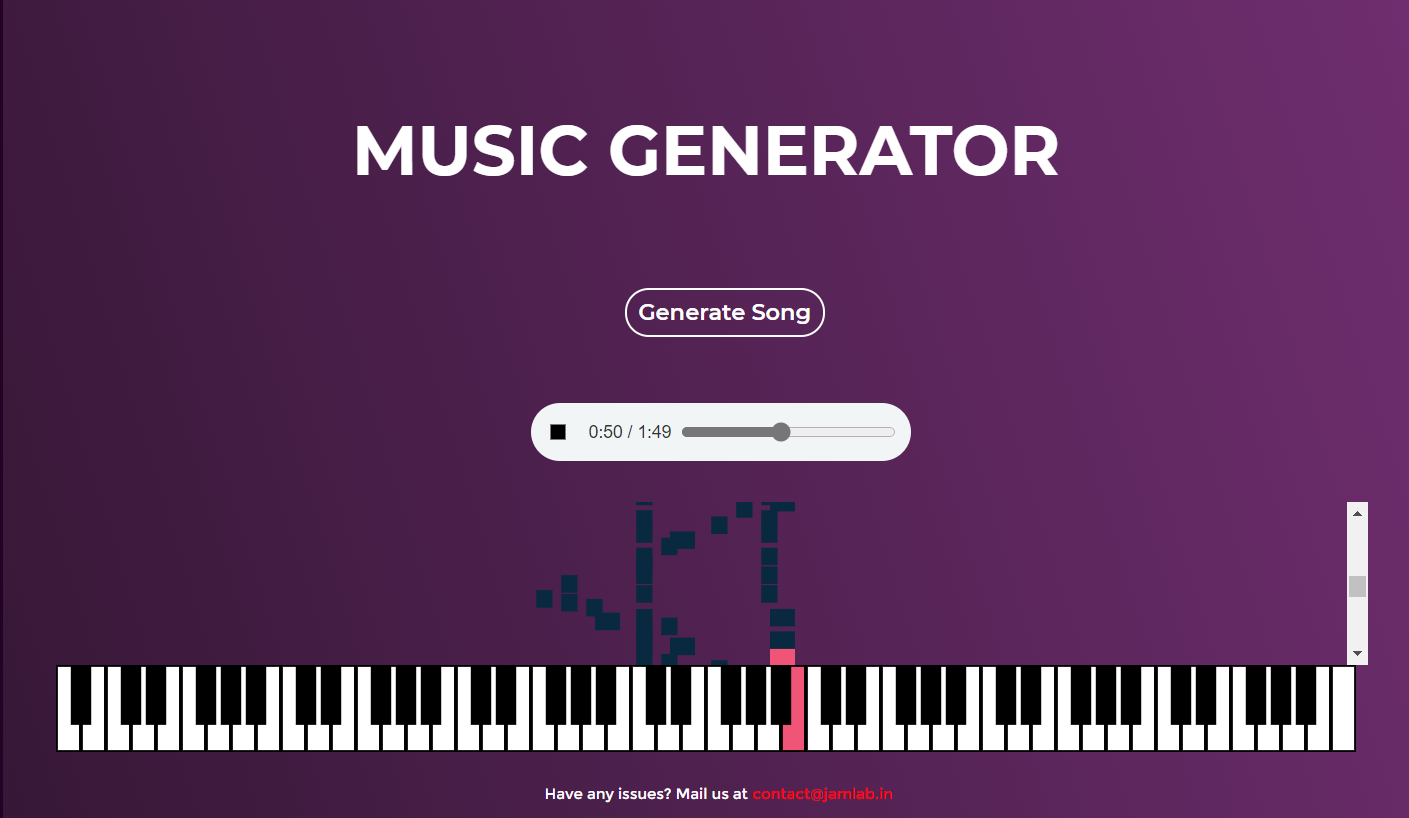}
\caption{Web Application}
\label{fig:12}
\end{figure}

Further, for better availability, the entire system has been converted into a web app. It can be reached at \url{https://research.jamlab.in/music-generator} . The web app has a highly intuitive interface as can be seen in Fig.\ref{fig:12}. There is a button which says "Generate Song". Upon pressing the button, the model will be invoked and a midi file will be generated. The generated midi file can be listened to using the midi player on the page itself. Once a file is played, it is visualized with a virtual piano displaying what keys are pressed to play the song.

\begin{figure}
\centering
\captionsetup{justification=centering}
\includegraphics[width=\textwidth,height=\textheight,keepaspectratio]{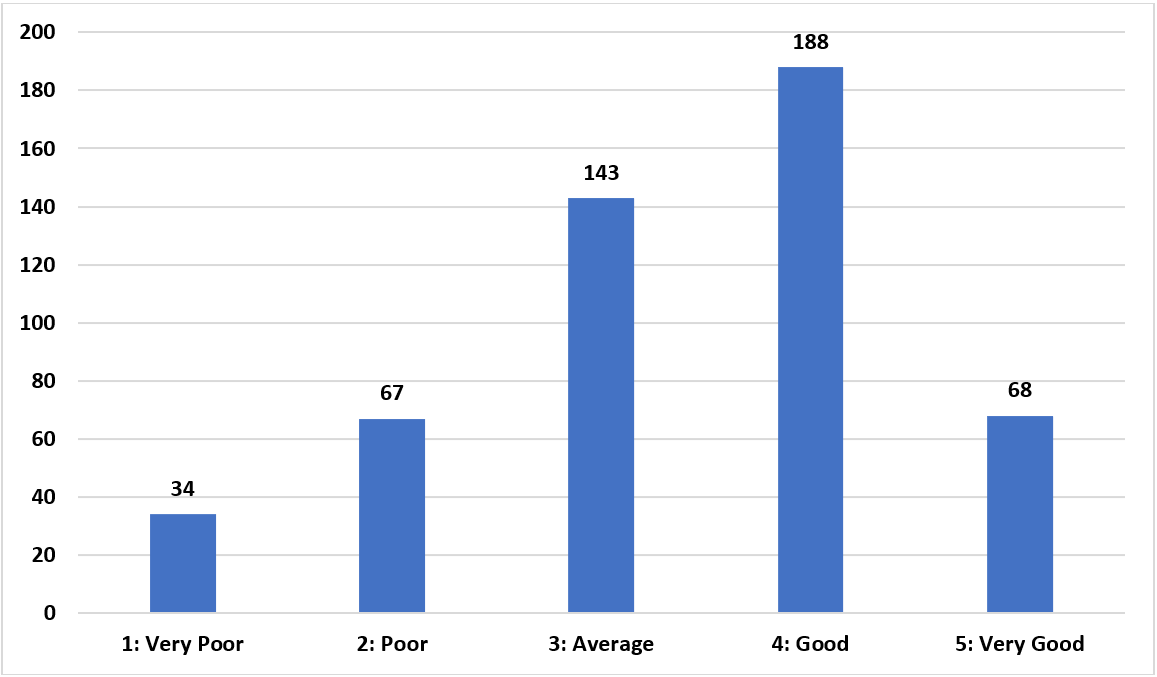}
\caption{MOS or Mean Opinion Score (Y-Axis shows the number of songs generated)}
\label{fig:13}
\end{figure}

The performance of this model was evaluated using a mean opinion score (MOS) test. 100 subjects were chosen for the survey, and 5 random songs, generated by the model, were played for each of them. The subjects were asked to rate the songs on a five-point scale (1:Very Poor, 2:Poor, 3:Average, 4:Good, 5:Very Good). A total of 500 different songs were evaluated this way, and a mean of the scores was evaluated. This MOS value came out to be 3.4. The survey has been summarized in Fig.\ref{fig:13}.

Fig.\ref{fig:13} suggests that most people think that the generated music is good. It can also be seen that many people think that it was average, but the generated compositions may instead be used as inspiration for new music.

\section{Conclusion}
A deep learning model capable of generating specific type of music by learning from MIDI data was created. The music is unique, and also pleasant, and overall the scores produced by the model have a good opinion value. Hence, it is possible to generate aesthetically pleasing music using a high level representation such as MIDI.

However, it can only generate a specific style/genre/mood of music that it has been trained on. These parameters, and using a larger and more comprehensive data set will be explored in future work.

\section{Future Work}
A lot of improvements can be made to the proposed model. This includes ability to generate specific mood/style/genre of music, training on a much larger dataset, and also exploring possibility of having different instruments and multiple tracks.

The piano roll based approach is also something that can be explored. Possibility of having lyric generation and singing synthesis could lead to a fully automated music production system.

%
%
%

\end{document}